\documentclass[sigconf,nonacm]{acmart}

\AtBeginDocument{%
  }

\setcopyright{acmcopyright}
\copyrightyear{2023}
\acmYear{2023}
\acmDOI{XXXXXXX.XXXXXXX}

\acmConference[FPGA '23]{ACM/SIGDA International Symposium on Field-Programmable Gate
   Arrays}{Feb. 12--14, 2023}{Monterey, CA}
\acmPrice{15.00}
\acmISBN{978-1-4503-XXXX-X/18/06}

\begin{document}

\title{AoCStream: All-on-Chip CNN Accelerator With Stream-Based Line-Buffer Architecture}

\author{Hyeong-Ju Kang}
\affiliation{%
  \institution{Korea University of Technology and Education}
  \streetaddress{1600 Chungjeol-ro}
  \city{Cheonan}
  \state{Chungcheongname-do}
  \country{Republic of Korea}
  \postcode{31253}
}
\email{hjkang@koreatech.ac.kr}

\renewcommand{\shortauthors}{Trovato et al.}

\begin{abstract}
Convolutional neural network (CNN) accelerators are being widely used
	for their efficiency, but they require a large amount of memory,
	leading to the use of a slow and power consuming external memory.
This paper exploits two schemes to reduce the required memory amount
	and ultimately to implement a CNN of reasonable performance 
		only with on-chip memory of a practical device like a low-end FPGA.
To reduce the memory amount of the intermediate data,
	a stream-based line-buffer architecture and a dataflow for the architecture
	are proposed instead of the conventional frame-based architecture,
		where the amount of the intermediate data memory
		is proportional to the square of the input image size.
The architecture consists of layer-dedicated blocks
	operating in a pipelined way with the input and output streams.
Each convolutional layer block has a line buffer
	storing just a few rows of input data.
The sizes of the line buffers are proportional to the width of the input image,
	so the architecture requires less intermediate data storage
		than the conventional frame-based architecture,
	especially in the trend of getting larger input size
		in modern object detection CNNs.
In addition to the reduced intermediate data storage,
	the weight memory is reduced by the accelerator-aware pruning.
The experimental results show that a whole object detection CNN can be
	implemented even on a low-end FPGA without an external memory.
Compared to previous accelerators with similar object detection accuracy,
	the proposed accelerator reaches much higher throughput
		even with less FPGA resources of LUTs, registers, and DSPs,
	showing much higher efficiency.
The trained models and implemented bit files are available at
	https://github.com/HyeongjuKang/accelerator-aware-pruning
	and https://github.com/HyeongjuKang/aocstream.


\end{abstract}



\maketitle

\newenvironment{processtable}[3]{
	\caption{#1}
	#2		
	\centering
	#3
}{}
\newcommand{\reffig}[1]{Figure~\ref{#1}}
\newcommand{\reftab}[1]{Table~\ref{#1}}
\newcommand{\refeqn}[1]{(\ref{#1})}

\def\form{2}
\section{Introduction}
Recently, convolutional neural networks (CNNs) are showing
	great performances in computer vision tasks, including
	image recognition
		\cite{he16, howard17, sandler18, tan19},
	object detection
		\cite{liu16_2, lin17, redmon18, tan20},
	and image segmentation \cite{long15}.
However, CNNs usually require an enormous amount of memory and computation,
	so special hardware is usually adopted to implement them.
Many kinds of CNN hardwares have been used,
	and a CNN accelerator in ASIC or FPGA shows high efficiency.

There have been many CNN accelerators
	\cite{zhang15, jo18, bai18, moon19, wen20},
	some of which focused on object detection
	\cite{fan18, wu19, nguyen21, lu21, pestana21, anupreetham21}.
One of the main concerns in designing a CNN accelerator
	is how to reduce the number of external memory accesses.
The processing of a CNN requires a large amount of memory,
	so the data are usually stored in an external memory like DRAM.
Accessing an external DRAM consumes much power \cite{han16}
	and occupies long latency.
Many previous CNN accelerators have proposed various data flows 
	to reduce the number of external memory accesses.

To solve this problem
	and ultimately to implement a whole CNN model only with on-chip memory,
	there are two trivial solutions, 
	embedding a large amount of on-chip memory \cite{jiao20, yuan21, meng21}
	or using a very simplified CNN model \cite{meng21}.
However, these solutions are not so practical
	because of cost and degraded performance.
It is still a challenging problem to reduce the required memory amount
	so that a CNN model of reasonable performance
	frequently used for embedded implementation \cite{xilinxmodelzoo, edgetpumodel}
	can fit in the on-chip memory
	of a practical environment like a low-end or mid-range FPGA device.

In this paper, we reached this goal by adopting two approaches.
The CNN processing stores two kinds of data in memory,
	the weights and the intermediate activation data.
To reduce the amount of weight memory, we exploit the pruning scheme
	\cite{han16, boo17, pang20, li22},
	especially accelerator-aware pruning \cite{kang19}.
Pruning schemes can reduce the weight amount, but the irregularity
	leads to an inefficient implementation.
The accelerator-aware pruning prunes weights considering the base accelerator,
	so it does not harm the accelerator performance.

To reduce the amount of the intermediate data memory,
	this paper proposes a stream-based line-buffer architecture.
The main component of a CNN is a convolutional layer.
The proposed architecture is specialized to process a convolutional layer,
	storing only a few rows of the intermediate data for each layer.
A convolution is a local operation, so the calculation of an output
	activation requires only a few neighboring input data.
If the input data are streamed into the processing block,
	only a few rows are required to be stored.
To take full advantage of the line-buffer structure,
	a proper dataflow will be proposed, too.
With the two schemes reducing the weight and the intermediate data memory,
	an object detection CNN can be implemented in a low-end FPGA
	without an external memory.

This paper is organized as follows.
Section II introduces the basics of CNN computations,
	and Section III analyzes the memory sizes of CNN accelerators.
The proposed architecture is described in Section IV,
	and the experimental results are shown in Section V.
Section VI makes the concluding remarks.

\section{Convolutional Neural Networks}
A CNN consists of many layers, which are stacked input-to-output.
The data usually flow from input to output.
The main layer in a CNN is a convolutional layer.
A convolutional layer assumes $N$ input feature maps
	whose height and width are $H$ and $W$.
A convolutional layer performs a convolution operation
	on the input feature maps as described in the following equation
	and produces $M$ output feature maps, as follows.
\begin{align}
fo(m,y,x) = \sum^{N-1}_{n=0} \sum^{K-1}_{i=0} \sum^{K-1}_{j=0}
				w(m,n,i,j) \times\\
			fi(n,S \times y + i, S \times x + j) + bias(m),	\nonumber
\label{eqn:conv}
\end{align}
where $fi()$ and $fo()$ are 
	a piece of the input and output feature map data, 
		an input and output activation, respectively,
	and $w()$ is the weights.

To reduce the amount of weight and computation,
	a convolution layer can be divided into a depth-wise convolution
		and a point-wise convolution \cite{howard17},
	where a point-wise convolution is a normal 1$\times$1 convolution.
In the depth-wise convolution, the number of the input feature maps, $N$,
	is equal to that of the output feature maps, $M$,
	and an output feature map is calculated
		from the corresponding input feature map.
\begin{align}
fo(n,y,x) = \sum^{K-1}_{i=0} \sum^{K-1}_{j=0}
				w(n,i,j) \times\\
			fi(n,S \times y + i, S \times x + j) + bias(n)	\nonumber
\label{eqn:convdw}
\end{align}

CNNs are usually used for computer vision tasks including object detection.
One of the most popular CNN types for object detection
	is the single-shot multi-box detector (SSD) \cite{liu16_2}.
The SSD exploits an image classification CNN like VGG, 
	ResNet, and MobileNet as a base CNN.
The feature maps shrink more with auxiliary layers,
	and the detection box information is generated through a few more layers.
There are some SSD variants, and SSDLite \cite{sandler18} uses 
	depth-wise convolutional layers instead of normal convolutional layers
	in the auxiliary part.

\section{Memory Size of CNN accelerators}
One of the most important factors in designing CNN accelerators
	is the amount of required memories.
Processing a neural network usually requires a huge amount of memories,
	usually larger than the amount that can be embedded
		on a low-end or mid-range FPGA.
A CNN accelerator, therefore, usually uses external memories like DRAMs.

A CNN accelerator stores two types of data in memories,
	weights and intermediate activations.
The amount of the weight memory is determined at the algorithm level by the CNN structure.
The amount of the activation memory is
	also determined at the algorithm level,
	but it can be determined at the architecture level, too.

Traditionally, the memory amount for weights is believed
	to be much larger than that for the activations.
In the traditional CNNs, however, most of the weights
	belong to the fully-connected layers
		\cite{han16}.
The recent CNNs use only one or none fully-connected layers
	\cite{he16, sandler18}, and 
	the object detection CNNs do not use fully-connected layers at all
	\cite{lin17, liu16_2, redmon18}.
In convolutional layers, the memory requirement for weights
	is not much larger than that for activations,
	compared to those in fully-connected layers.

Furthermore, the activation amount is proportional
	to the square of the input image size.
If the height and width of the input image are doubled,
	so are those of feature maps,
	and the activation amount increases by four-times.
This is not a big problem when the target is the image classification
	because the input image size is usually very small, around 224.
However, the modern object detection CNNs use large input images
	varying from 300 \cite{liu16_2} to 1280 \cite{tan20}.
Considering the current trend of processing larger input images,
	the activation memory will become larger in the future.

The amount of the activation memory also depends
	on the accelerator architecture.
In the conventional CNN accelerators,
	a neural network is processed layer by layer.
A whole input feature map is stored in a memory,
	and a CNN accelerator reads activations from the memory, processes them,
	and stores the output activations.
After generating the whole output feature maps,
	the CNN accelerator starts to process the next layer.
Therefore, the CNN accelerator requires a memory
	for the whole input or output feature maps,
	and the amount is sometimes doubled for the double buffering.
Some structures process a few layers at the same time \cite{alwani16, bai18},
	but they store the intermediate data between the layer blocks, too.
\begin{figure}[!t]
	\centering
	\includegraphics[scale=0.70]{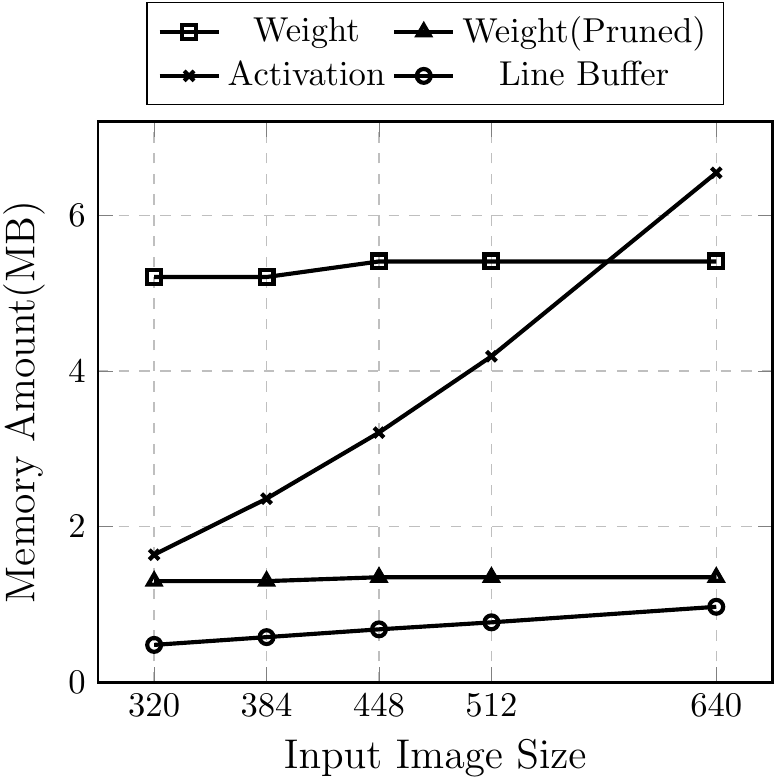}
	\caption{Memory amount for MobileNetV1 and SSDLiteX}
	\Description{The memory amounts are analyzed for various input image sizes.}
	\label{fig:resource}%
\end{figure}

To analyze the memory size, a few object detection CNNs were designed.
The CNNs consist of MobileNetV1 \cite{howard17}
	and SSDLiteX, a variant of SSDLite \cite{sandler18}.
The CNNs are built for the images with various sizes from 320 to 640.
The number of auxiliary layer stages changes with the input sizes.
The input size 320 and 384 uses 4 stages, 448 and 512 uses 5 stages,
	and 640 uses 6 stages.
The detailed CNN structure is provided in \cite{kangmodel}.
\reffig{fig:resource} compares the memory amounts for each type of data
	with 8-bit quantization.
For small input images, the memory amount for the activations
	(\textit{Activation} in \reffig{fig:resource})
	is around one-fourth of that for the weights
	(\textit{Weight} in \reffig{fig:resource}).
With large input images, the activations occupy almost the same memory
	as the weights do.
If the double buffering scheme is applied,
	the intermediate activations require twice as large memories
		as that in the figure.
\begin{figure}[!t]
	\centering
	\includegraphics[scale=0.70]{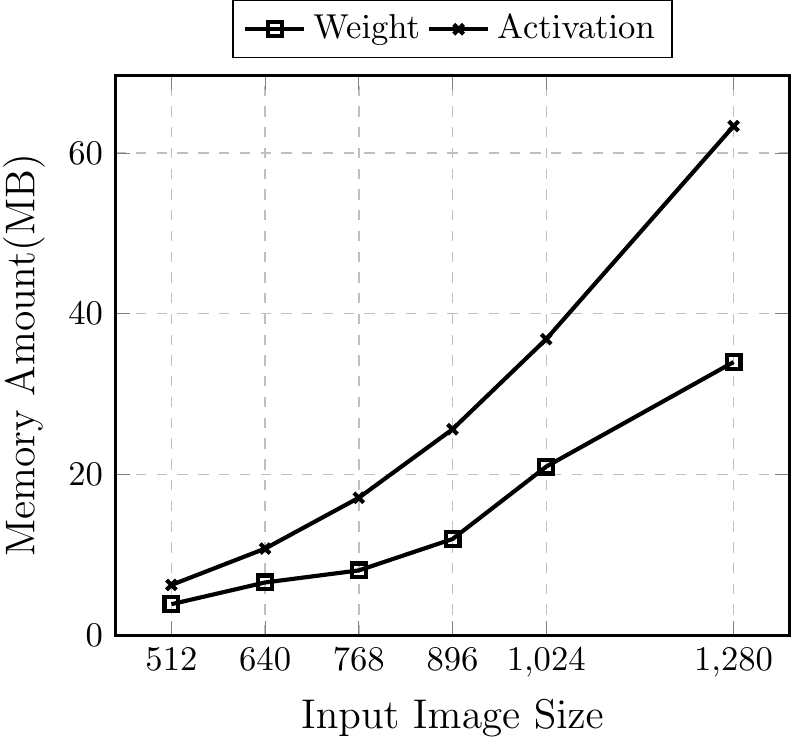}
	\caption{Memory amount for EfficientDet}
	\Description{The memory amounts of EfficientDet are analyzed for various input image sizes.}
	\label{fig:res_effdet}%
\end{figure}

As another example of the state-of-the-art object detection CNNs,
	\reffig{fig:res_effdet} shows the memory requirement
	of EfficientDet \cite{tan20}, where
		the number of channels increases according to the input size.
The amount of weights increases as input size is scaling up,
but the amount of the activation memory is larger than that of the weight memory.

Furthermore, the weight amount can be reduced by pruning
	\cite{han16, boo17, pang20, li22, kang19}.
Recent research on pruning shows the amount of weight
	can be reduced by three fourths in convolutional
		layers \cite{han16, kang19}. 
If the pruning is applied, the memory amount for weights is smaller than
	that for activations even with small input images,
	as shown in \reffig{fig:resource}.
The pruning can reduce the weight amount, 
	but there is no method to reduce the activation amount.
The only way is using a smaller input image
	despite of the performance degradation or using another architecture.

\section{Stream-Based Line-Buffer Architecture}
This paper follows the accelerator-aware pruning
	and the corresponding PE structure in \cite{kang19}
		to reduce the weight amount,
	so the remaining part will focus on the reduction of the activation memory
		in the architecture and dataflow level,
	proposing a stream-based line-buffer accelerator architecture for CNNs.

\subsection{Top architecture}
\begin{figure}[!t]
	\centering
	\includegraphics[scale=0.55]{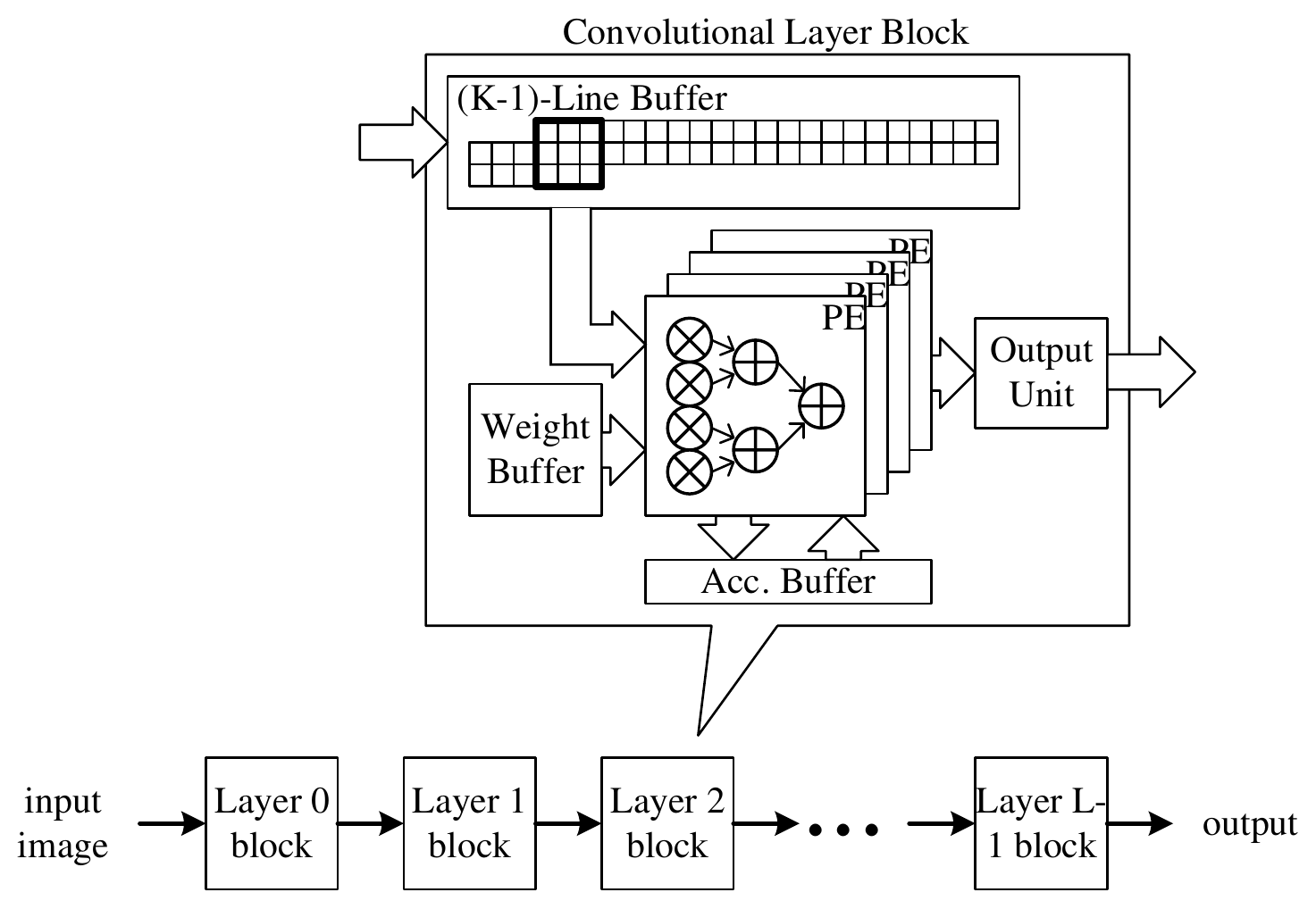}
	\caption{Stream-based line-buffer architecture}
	\Description{The detailed block diagram of the proposed architecture.}
	\label{fig:top_arch}%
\end{figure}

The proposed accelerator processes a CNN in a layer-level pipelined way.
Each layer has a corresponding processing block
	as shown in \reffig{fig:top_arch}.
When a group of data is input to a block, the block processes the input data
	and generates a group of output data if possible.
The generated group of data streams into the next block.
Since each block does not wait for the previous block
	to complete the corresponding layer operation,
	all of the blocks can operate in parallel.
The structure of a layer block is determined by the corresponding layer type.

\subsection{Convolutional layer block}
The base operation of a convolutional layer is the two-dimensional convolution.
In the conventional image processing circuits, the two-dimensional convolution
	is usually processed by a stream-based structure
		with a line buffer of size $K$-1 lines.
In the structure, the input data is assumed not to reside in a memory,
	but to stream in one by one.
When one piece of input data streams in, the circuit
	processes the possible convolution operation.

\begin{figure*}[!t]
	\centering
	\includegraphics[scale=0.65]{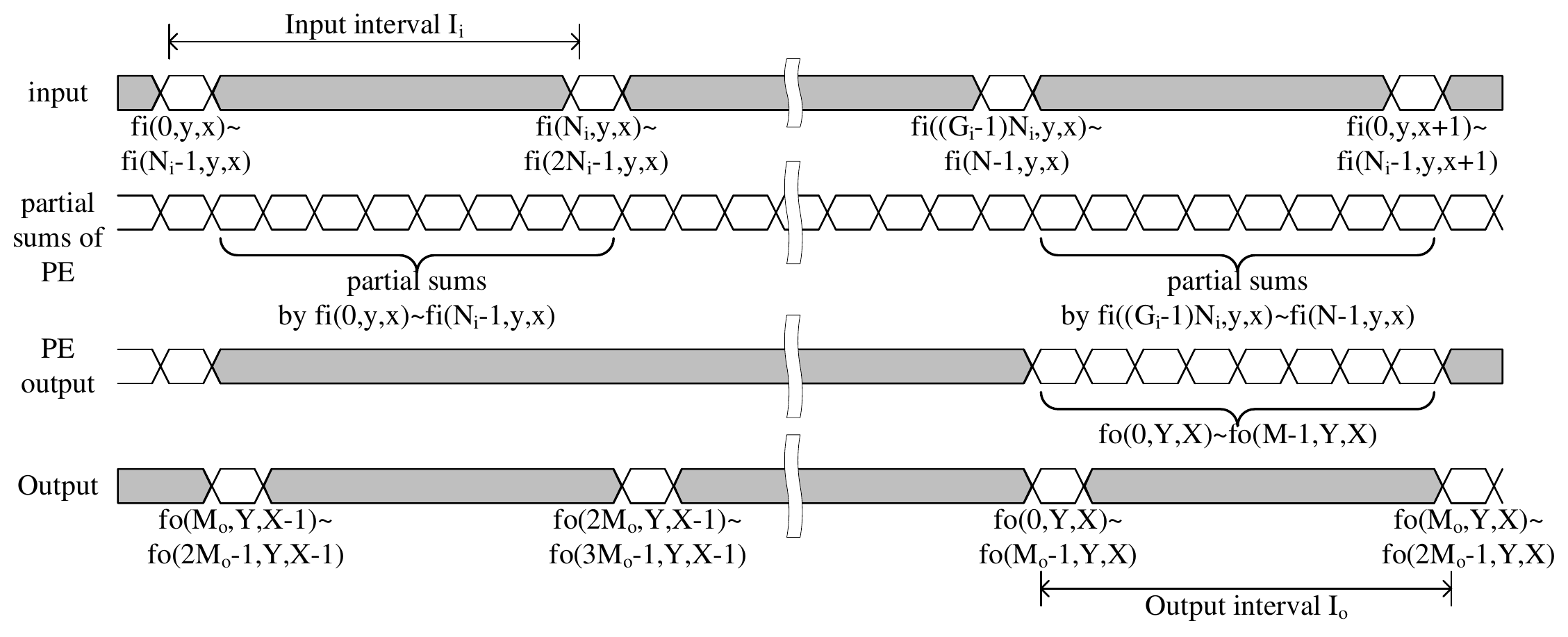}
	\caption{Convolutional layer processing}
	\Description{The timing diagram of convolution processing in the proposed architecture.}
	\label{fig:conv_time}%
\end{figure*}

The stream-based structure can be applied to the convolutional layer operation,
	but a proper dataflow is required
		to maintain the $K$-1 line-buffer size.
As in the typical image processing circuits,
	it is assumed that the input data are streamed in the row-major order.
For each spatial location, $N$ channel data are divided into $G_i$ groups,
	and a group of $N/G_i = N_i$ data is streamed-in together
	at the interval of $I_i$ cycles as shown in \reffig{fig:conv_time}.
With the $N_i$ data, the layer block performs all of the computations
	that can be done with the input data and the data stored in the buffer.
When $g$th group data,
		$fi(gN_i,y,x) \sim fi((g+1)N_i-1,y,x)$, are input, 
	the layer block calculates the following partial sums for each output
		$fo(m,Y,X)$, where $0 \le m < M$, $Y = y-K+1$, and $X = x-K+1$.
For simplicity, the stride S is assumed to be 1,
	but the structure is not limited to that.
\begin{align}
fo_g(m,Y,X) = 
		\sum^{(g+1)N_i - 1}_{n=g N_i} \sum^{K-1}_{i=0} \sum^{K-1}_{j=0}
				 w(m,n,i,j)	\\
		\times fi(n,Y+i,X+j),	\nonumber
\label{eqn:conv_part}
\end{align}

The partial sum requires $K \times K \times N_i \times M \times (1-r)$
	MAC operations, where $r$ is the pruning ratio,
	and the operations should be done in $I_i$ cycles.
Therefore, the required number of MAC operators is
	$K \times K \times N_i  \times M_i \times (1-r)$, where $M_i = M / I_i$.
The layer block has $M_i$ processing elements (PEs),
	and a PE calculates a partial sum of an output
		with $K \times K \times N_i \times (1-r)$ multipliers at each cycle.

When a partial sum is calculated, it is accumulated with an accumulation buffer 
	of size $M$.
When all the data of a spatial location,
	$fi(n,y,x)$ for $0 \le n < N$,
	are input through $G_i$ groups,
	the calculation of the output data, $fo(m,Y,X)$ for $0 \le m < M$,
	is completed through the accumulation.
The output data are collected at the output unit
	and streamed out in $G_o$ groups of $M/G_o = M_o$ data
	at the interval of $I_o$ cycles.
If the spatial size of the input feature maps is equal
	to that of the output feature maps,
	the following relationship should be satisfied.
\begin{equation}
	\frac{N}{N_i} \times I_i \ge \frac{M}{M_o} \times I_o
\end{equation} 

Some similar line-buffer structures were proposed for CNNs
	in \cite{zhang18_2, blott18, nguyen21, meng21, aarrestad21, anupreetham21}.
However, they did not employ a dataflow proper to the line-buffer structure.
Their dataflow focuses on the weight data reuse,
	leading to the larger line-buffer of size $K$ or $K$+1 lines.
The large line-buffers make their accelerators
	use an external memory for the weights \cite{zhang18_2, nguyen21}
	or require a very large FPGA device \cite{meng21, anupreetham21}.

On the contrary to the previous works,
	the proposed dataflow reuses the input feature map data as much as possible.
After a $K \times K \times N_i$ input activation data block is gathered,
	the PEs perform all the computations related to the block.
This dataflow property enables the line-buffer size of $K$-1 lines.
However, this dataflow cannot reuse the weights,
	so it is proper to a structure with all of the weights in on-chip memory.
With a weight memory reduction scheme,
	the proposed structure can exploit the dataflow 
		to reduce the line-buffer size.


\subsection{Depth-wise convolutional layer block}
The depth-wise convolutional layer block also requires a line buffer
	of ($K$-1)-line size as the convolutional layer block
		in the previous subsection.
In the depth-wise convolution, the accumulation is not required between
	the input data groups.
When $fi(gN_i,y,x) \sim fi((g+1)N_i-1,y,x)$ data are input,
	we can calculate $fo(gN_i,Y,X) \sim fo((g+1)N_i-1,Y,X)$.
The required number of MAC operations is $K \times K \times N_i$.
Each PE has one MAC unit and the number of PEs is determined as follows.

\begin{equation}
	\text{Number of PEs} \ge \frac{K \times K \times N_i}{I_i}
\end{equation}

\subsection{Frame-buffer vs. line-buffer}
When a CNN is processed layer-by-layer as in the conventional architecture,
	a frame buffer is required.
For a layer $l$, 
	a frame buffer of size $H_l \times W_l \times N_l$
	is required for the input feature maps,
	and another of size $H_{l+1} \times W_{l+1} \times N_{l+1}$
	is required for the output feature maps.
Since a frame buffer can be reused between layers,
	the maximum size is required as follows.

\begin{equation}
	\text{Frame Buffer Size} = \max_{l} H_l \times W_l \times N_l
	\label{eqn:frame_size}
\end{equation}

In the proposed architecture, a line buffer is used for each
	convolutional layer block, depth-wise convolutional layer block,
	and pooling layer block.
Since the blocks operate in parallel, the line buffers cannot be shared.
Therefore, the total size of the line buffers is as follows.

\begin{equation}
	\text{Line Buffer Size} = \sum_{l} (K_l-1) \times W_l \times N_l
	\label{eqn:line_size}
\end{equation}

When the input image size is scaled-up, the input image is enlarged
	vertically and horizontally.
The frame buffer size in \refeqn{eqn:frame_size} is increased
	with the square of the scale.
Contrary to that, the line buffer size in \refeqn{eqn:line_size}
	has only the width term, $W_l$.
The line buffer size is proportional to the scale linearly.
In \reffig{fig:resource}, the frame buffer size is shown as \textit{Activation},
	which increases rapidly with the input image size.
However, the line buffer size, denoted as \textit{Line Buffer},
	increases slowly with the input image size. 

\begin{figure}[!t]
	\centering
	\includegraphics[scale=0.70]{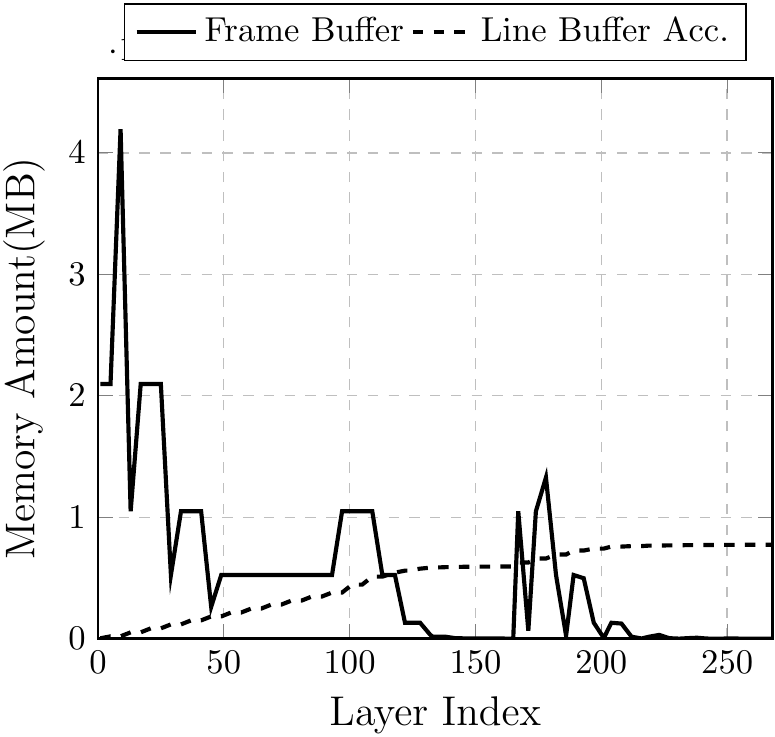}
	\caption{Frame-buffer vs. line-buffer}
	\Description{The memory usage is compared for the frame-buffer architecture and the proposed architecture.}
	\label{fig:buf_size}%
\end{figure}

\begin{table*}[!t]
	\processtable{FPGA Implementation Results}
	{\label{tab:fpga}}
	{
\begin{tabular}{ l | r r r | r r r r | r | r}
	\toprule
		Architecture & \cite{wu19} & \cite{fan18} & \cite{nguyen21} 
					 & \multicolumn{5}{c}{AoCStream (Proposed)}	\\
	\midrule
		CNN		 & MNetV1+SSD & MNetV2+SSDLite & YOLOv3
				 & \multicolumn{6}{c}{MNetV1 + SSDLiteX}	\\
		Input Size		& 320  & 224  & 416  & 320  & 384  & 448  & 512  & 320  & 448  \\
		MS COCO AP		& .193 & .203 & .310 & .211 & .231 & .247 & .253 & .206 & .247 \\
	\midrule
		FPGA		 & XCZU9EG & ZC706 & XC7VX485T
					 & \multicolumn{4}{c|}{XCKU5P} & XC7K325T & XC7VX485T \\
		LUT(K)			& 162   & 148  & 230   & 137  & 145  & 148  & 154  & 155  & 154  \\
		Reg(K)			& 301   & 192  & 223   & 218  & 219  & 233  & 232  & 195  & 237  \\
		BRAM			& 771   & 311  & 972.5 & 454  & 454  & 476  & 476  & 445  & 648  \\
		URAM			& -     & -    & -     & 25   & 25   & 25   & 44   & -    & -    \\
		DSP     		& 2070  & 728  & 2640  & 464  & 464  & 476  & 476  & 360  & 476  \\
		Clock(MHz)		& 333   & 100  & 200   & 428  & 349  & 400  & 375  & 186  & 223  \\
		Throughput(FPS)	& 124.3 & 15.4 & 11.66 & 260.9& 147.7& 124.5& 89.3 & 100.9& 69.5 \\
		DSP Efficiency 1(\%)&22.3& 8.3 & -     & 80.2 & 80.2 & 78.3 & 78.3 & 76.1 & 78.3 \\
		DSP Efficiency 2(\%)& - & -    & 72.4  & 289  & 289  & 282  & 282  & 332  & 282  \\
		Ext. Mem.  		& WA    & WA   & W     & None & None & None & None & None & None \\
	\bottomrule
\end{tabular}
	} {}
\end{table*}


\reffig{fig:buf_size} compares the frame buffer size and the accumulated line buffer
	in each layer of MobileNetV1 and SSDLiteX
		with 512$\times$512 input image.
The maximum size of the frame buffer is 4M
	at the output of the first point-wise convolution.
The line buffer in each layer is very small, so it would not be clearly shown
	in the figure.
Instead of the line buffer size in each layer,
	the figure illustrates the accumulated line buffer amount, 
	which is less than one fourth of the frame buffer size.
\begin{figure}[!t]
	\centering
	\includegraphics[scale=0.7]{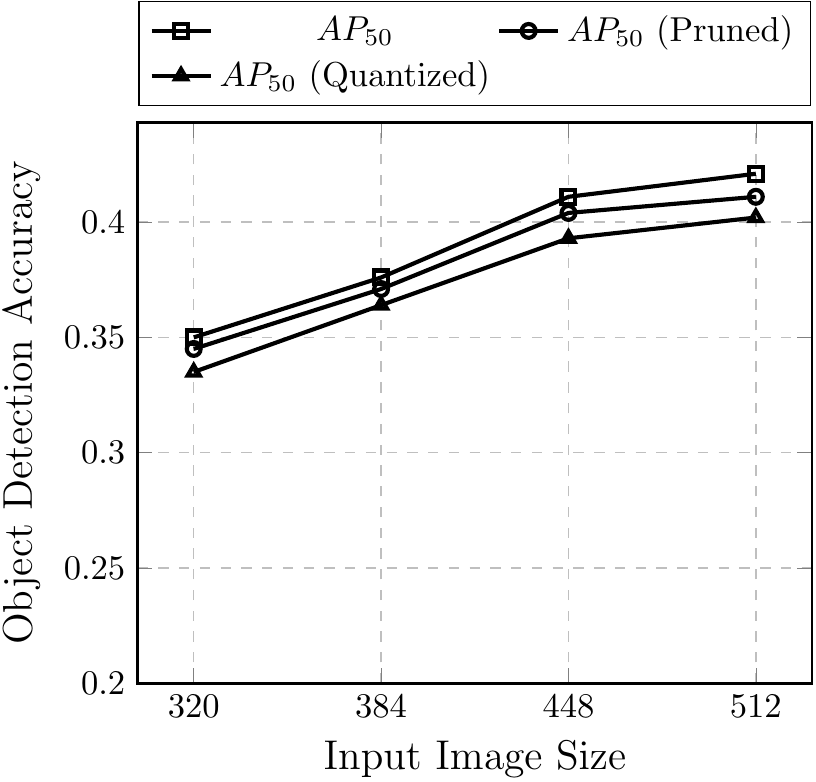}
	\caption{Object detection accuracy for MS COCO dataset}
	\Description{The accuracy of the CNN models used in the experiments.}
	\label{fig:accuracy}%
\end{figure}

\subsection{All-on-chip accelerator}
The weight pruning and the line buffer architecture
	reduces the storage of the weights and the intermediate data, respectively,
	so their combination can lead to all-on-chip implementation.
The two schemes can reduce the memory size by around three-fourths.
For example, the 512$\times$512 input image case in \reffig{fig:resource}
	requires the weight memory of around 5MB
	and the intermediate data memory of around 4MB.
The total memory requirement of 9MB cannot be afforded
	by a low-end or mid-range FPGA device
	like Xilinx XCKU5P, whose on-chip memory size is 4MB.
If the two schemes are applied, the total memory size becomes
	around 2.3MB, which is less than the on-chip memory size of XCKU5P.

The proposed scheme does not guarantee that any CNN can be implemented
	only with the on-chip memory of any device.
There will be no such scheme.
The proposed scheme, however, broadens the possibility of the all-on-chip implementation,
	higher performance CNNs on smaller devices.

\section{Experimental Results}
Object detection CNNs based on MobileNetV1 and SSDLiteX
	with various input sizes
	are trained and implemented with the proposed architecture.
The CNNs are trained with the MS COCO data set and
	pruned by the accelerator-aware pruning.
The pruning ratio is 75\%, which means 6 weights are pruned
	for every 8 weights along the channel axis.
The pruned CNNs are quantized with 8--10 bits without fine-tuning.
The object detection accuracy, $AP_{50}$, for the MS COCO dataset
	is provided after each step of training, pruning, and quantization
	in \reffig{fig:accuracy}.
Pruning and quantization degrade $AP_{50}$ by around 0.01--0.02,
	but the detection accuracy is still high for such compact CNNs.
If retraining is applied with quantization,
	better detection accuracy could be obtained.
The final $AP$ values are shown at the fourth row of \reftab{tab:fpga}.

The proposed accelerator is designed in the register-transfer level (RTL)
	for the quantized CNNs
	and implemented for a low-end Xilinx FPGA, XCKU5P,
	which is the second smallest device in the UltraScale+ Kintex series.
The implementation results are shown on the right side of \reftab{tab:fpga}.
The table shows the occupancy of FPGA resources including
	look-up tables (LUT), registers, block memories (BRAM),
	ultra memories (URAM), and DSP units.
The last two columns are the implementation results for older FPGAs,
	Kintex-7 series and Virtex-7 series,
	for comparison with previous works.
Becasue of the resource limitation, 
	some layers are pruned to 87.5\% for XC7K325T.

The last four rows of the table show
	the operating clock frequency, the throughput in frames per second,
	the DSP efficiency, and the external memory use.
The DSP efficiency is calculated as follows.
\begin{align}
\mathrm{DSP\;Efficiency\;1} = \frac{\mathrm{(Operations/Frame)} \times
						\mathrm{(Frames/second)}}
			{2 \times \mathrm{(Num.\;of\;DSPs)} \times \mathrm{(Clock\;Freq.)}}
\label{eqn:dsp_eff}
\end{align}
, where the $2 \times$ in the denominator reflects that a DSP can process
	two operations, a multiplication and an addition, simultaneously.
The second DSP efficiency is the effective efficiency,
	which includes the zero-skipped operations in a sparsity architecture,
	so the effective efficiency can be higher than 100\% if pruning is applied.
At the last \textit{Ext. Mem.} row, \textit{W} and \textit{A}
	means the weights and the activations are
	stored in external memories, respectively.
As the table shows, the proposed architecture can store
	the whole intermediate data and weights on the on-chip BRAM and URAM
	even for the input image size 512$\times$512.
The all-on-chip implementation leads to high throughput and efficiency.
The architecture can process images in 90 to 250 fps,
	which is much faster than the real-time speed, 30fps.

The table also compares the proposed architecture with the previous ones.
The architectures of the second and third columns use
	CNNs similar to the one used in this paper.
The accelerator of \cite{wu19} used the MobileNetV1 and SSD combination.
Their architecture shows the highest throughput in the previous ones,
	but it is slower even with around five-times more DSPs
		than the proposed architecture of the same input size.
The architecture of the third column used MobileNetV2 and SSDLite
	with a small input size \cite{fan18}.
Despite of such small input size and high DSP usage,
	the throughput is very low.
The accelerators of \cite{wu19} and \cite{fan18} are based on the frame-based
	architecture, so their DSP efficiency is very low
		because of the DRAM accesses.
At the fourth column, an accelerator using YOLOv3
	is compared \cite{nguyen21}.
The accelerator exploits a line buffer architecture similar to the proposed one,
	but it uses a larger line buffer in each layer
	and still needs an external memory for weights.
Because they used a different CNN, it is difficult
	to compare their accelerator with the proposed one,
	but the DSP efficiency is lower than AoCStream.
Even though they adopted a sparsity architecture,
	the effective DSP efficiency is not higher than 100\%
	probably because of DRAM accesses for weights.
HPIPE proposed in \cite{anupreetham21} also used a line buffer architecture,
	but it is implemented on a different type of FPGA.
Because of the different internal FPGA structure,
	it is not compared in the table.
However, HPIPE uses much resource to implement a similar object detection CNN, 
	4,434 DSPs and 7,179 M20K BRAMs for MobileNet v1 and SSD,
	showing less-than-50\% DSP efficiency.

\section{Conclusion}
In this paper, object detection CNNs with reasonable performance
	were implemented only with the on-chip memory
	of a practical device.
The memory amount is reduced in the algorithm level, accelerator-aware pruning,
	and in the architecture level, a stream-based line buffer architecture.
In the architecture, 
	a dedicated block is assigned to each layer,
	and the layer blocks operate in a pipelined way.
The intermediate data are streamed into and out of each block,
	so only a few rows are stored in each block
		thanks to the property of the convolution operation.
The reduction of the intermediate data storage is combined
	with the reduction of the weight storage by pruning
	to remove the need of external memories.
The all-on-chip implementation greatly enhances the performance
	of the CNN accelerator.
The architecture can be applied to various CNNs for other computer
	vision tasks.


\bibliographystyle{../../Templates/ACM/acmart-primary/ACM-Reference-Format.bst}
\bibliography{../../../references/Machine_Learning/cite}


\end{document}